\documentclass[twocolumn,aps,prl,showpacs,preprintnumbers,superscriptaddress]{revtex4}
\usepackage{amssymb}
\usepackage{graphicx}

\begin{document}

\title{Conical emission, pulse splitting and X-wave parametric amplification
in nonlinear dynamics of ultrashort light pulses}

\author{Daniele Faccio}
\email{daniele.faccio@uninsubria.it} \affiliation{INFM and Department of Physics \& Mathematics, University
of Insubria, Via Valleggio 11, 22100 Como, Italy}
\author{Miguel A. Porras}
\affiliation{Departamento de Fis\'{\i}ca Aplicada, ETSIM, Universidad Polit\'ecnica de Madrid, Rios Rosas
21, 28003 Madrid, Spain}
\author{Audrius Dubietis}%\author{Algis Piskarskas}
\affiliation{Department of Quantum Electronics, Vilnius University, Sauletekio Ave. 9, 10222 Vilnius,
Lithuania}
\author{Francesca Bragheri}
\address{Department of Electronics, University of Pavia, Via Ferrata 1, 27100 Pavia, Italy}
\author{Arnaud Couairon}
\affiliation{Centre de Physique Th{\'e}orique, {\'E}cole Polytechnique, CNRS UMR 7644, 91128 Palaiseau
Cedex, France}
\author{Paolo Di Trapani}
\affiliation{INFM and Department of Physics \& Mathematics, University of Insubria, Via Valleggio 11, 22100
Como, Italy}

\begin{abstract}
The precise observation of the angle-frequency spectrum of light filaments in water
reveals a scenario incompatible with current models of conical emission (CE). Its
description in terms of linear X-wave modes leads us to understand filamentation
dynamics requiring a phase- and group-matched, Kerr-driven four-wave-mixing
process that involves two highly localized pumps and two X-waves. CE and temporal
splitting arise naturally as two manifestations of this process.
\end{abstract}


\maketitle

Filamentation of intense light pulses in nonlinear media has attracted much interest
ever since first experimental evidences in the early '60's
(\cite{marburger:1975} and references therein). Owing to the very high
intensities reached during the process, several nonlinear phenomena, e.g.,
multi-photon absorption, plasma formation, saturable nonlinear response, stimulated
Raman scattering etc., occur in addition to the optical Kerr effect. Indeed, the
filament regime is enriched by peculiar phenomena like pulse splitting,
self-steepening, shock-wave formation, super-continuum generation, and conical
emission (CE) \cite{gaeta:2003}. In media with normal group velocity dispersion (GVD), no matter if of
solid, liquid or gaseous nature, CE accompanies filamentation, producing
radiation at angles that increase with increasing detuning from the carrier
frequency \cite{nibbering:1996,faccioJOSAB:2005}. In spite of the generality of the process, a clear understanding of the
interplay between CE and filament dynamics is still missing. Only recently, Kolesik
{\em et al.} have proposed an interpretation of filamentation dynamics in water on
the basis of pulse splitting and dynamic nonlinear X waves at the far field
\cite{kolesikprl:2004}, in which the double X-like structure observed in simulated
angle-frequency spectra arises from the scattering of an incident field at the two
main peaks of the split material response wave.

%The concept of nonlinear X-wave (at the near and far fields) was originally introduced for interpreting
%spontaneous spatiotemporal localization in second-harmonic generation. These are weakly localized,
%stationary, nonlinear phase modulations and group velocity are balanced with angular dispersion, i.e. with
%dependence of propagation angle with frequency.

%\cite{ditrapani:2003}. Nonlinear X-waves inherently carry AD, more precisely, they carry the same
%angular dispersion amount as linear X-waves \cite{contipre:2004}. These considerations justify an
%attempt of characterizing CE appearing in conjunction with the NL filamentation in terms of \emph {linear X
%waves}, which is the subject of this work. %%

Originally, CE in light filaments was interpreted in terms of the modulation
instability (MI) angle-frequency gain pattern of the plane and monochromatic (PM)
modes of the nonlinear Schr\"{o}dinger equation (NSE)
\cite{luther2:1994,mckinstrie:1992}. Measurements at large angles and detunings from
the carrier frequency gave in fact results fairly compatible with this interpretation
\cite{alfano:1970,alfano:1993}. In the present work, owing to the use of a novel
imaging spectrograph technique \cite{faccioJOSAB:2005}, we have been able to observe
for the first time the CE in the region of small angles and detunings. The results
clearly indicate a scenario not compatible with the MI analysis of PM modes. Our
description by means of the spectra of the stationary {\em linear X-waves} supported
by the medium, indicates that the strong localization of the self-focused field plays
a crucial role in the substantial modification experienced by the MI pattern. We
propose a simple picture in which the latter results from the parametric
amplification of two weak X-waves by the strong, highly localized pump. Supporting
this interpretation, we are able to derive, from the matching condition among the
interacting waves, a simple analytical expression [Eq. (\ref{betaI})] that accurately
determines the overall CE structure, and predicts also the splitting velocity of the
pump wave as a function of the peak intensity reached at collapse.

%These results, being consequence of the cubic nonlinear Schr\"odinger dynamics, can
%be readily translated to understand the post-collapse dynamics of the different
%nonlinear waves commonly described in the frame of the NSE, as Bose-Einstein
%condensates or Langmuir plasma waves.

\begin{figure}
\includegraphics[angle=-90,width=7.5cm]{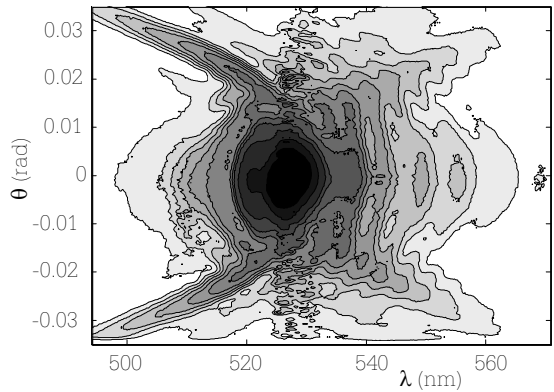}
\caption{\label{fig:fig1} Measured $\theta$--$\lambda$ spectrum for $E_{\rm in}=2$ $\mu$J.}
\end{figure}

The experiment was carried out using a 3 cm long cell filled with water as a bulk
nonlinear Kerr medium. Beam filamentation was induced by launching 200 fs pulses at
carrier wave length 527 nm, delivered from a mode-locked, chirped-pulse
regeneratively amplified, Nd:glass laser with a 10 Hz repetition rate (Twinkle; Light
Conversion Ltd.). The pulse energy was controlled by use of a half-wave plate and a
Glan-Taylor polarizer. A spatial filter guarantees high beam quality before focusing
with a $f = 50$ cm lens placed at 48 cm from the cell input facet, with a beam
diameter at half-maximum equal to 100 $\mu$m. The output facet is then imaged with a
4-$f$ telescope onto the rear focal plane of the lens ($f_{F}$ = 15 cm) used to
obtain the spatial Fourier transform of the filament generated in the Kerr sample.
The input slit of the imaging spectrograph (MS260i, Lot-Oriel) with a high resolution
1200 lines/mm diffraction grating is placed at a distance $f_{F}$ from the Fourier
lens so as to reconstruct the angle-wavelength ($\theta$--$\lambda$) spectrum of the
filament on a commercial CCD 8-bit camera (Canon), placed at the output plane of the
spectrograph. With the change $\lambda = 2\pi c/(\omega_0+\Omega)$, the
angle-frequency ($\theta$--$\Omega$) spectrum can also be obtained, $\Omega$ being
the detuning from the carrier frequency $\omega_0$. More details of the experimental
layout may be found elsewhere \cite{faccioJOSAB:2005}. We recorded only single-shot
spectra in order to avoid averaging effects due to possible shot-to-shot fluctuations
in the input pulse energy. Moreover we highly saturated the central peak of the
spectrum so as to highlight the less visible surrounding structure. For input
energies $E_{\rm in}\lesssim 1.8\,\mu$J, no CE, or clear X-like features were
observed. Figure \ref{fig:fig1} shows an example of $\theta$--$\lambda$ spectrum at
$E_{\rm in}=2\,\mu$J. The CE appears as distinctly separated red- and blue-shifted
X-shaped tails. This pattern remains very similar with increasing input energy up to
$ E_{\rm in}\sim 4$ $\mu$J, while further increase produces a slowly deteriorating
picture, with a modulated intensity pattern that extends to nearly all recorded
values of $\theta$ and $\lambda$. We interpret this deterioration as due to the onset
of local breakdown in the water sample.

CE emission is commonly accepted to arise from the interplay of diffraction, dispersion and nonlinear
material response, the simplest model that accounts for it being the cubic NSE with normal GVD
\cite{luther2:1994,mckinstrie:1992}
\begin{eqnarray}\label{eq:NLSE}
\frac{\partial A}{\partial z}= \frac{i}{2k_0}\nabla_{\bot}^{2}A -
\frac{ik_0^{\prime\prime}}{2}\frac{\partial^{2}A}{\partial \tau^{2}} + i \frac{\omega_0 n_{2}}{c}|A|^{2}A
\, .
\end{eqnarray}
Here, $A(x,y,\tau,z)$ is the complex envelope of the wave packet $E = A \exp(ik_0 z - i\omega_0 t)$ of
carrier frequency $\omega_0$, $\nabla^2_\bot = \partial^2/\partial x^2 +
\partial^2/\partial y^2$ is the two-dimensional Laplace operator perpendicular to the
propagation direction $z$, $\tau = t- k'_0z$ is the local time, $c$ the speed of light in vacuum, $n_2$
the nonlinear refractive index, and $k_0^{(m)} \equiv [d^{m} k(\omega) /d\omega^m]_{\omega_0}$, where
is $k(\omega)$ the frequency-dependent propagation constant. Mathematically, the $\theta$--$\Omega$ spectrum
where CE is observable, is directly related to the Fourier spectrum of the envelope. If for instance,
$A(r,\tau,z)$ [$r =(x^2+y^2)^{1/2}$] is a cylindrical symmetric complex envelope, and
$\hat A(K_\perp,\Omega,z)$ [$K_\perp =(k_x^2+k_y^2)^{1/2}]$ is its spatiotemporal Fourier transform, then
the $\theta$--$\Omega$ spectrum is given by $\hat A(k_0\theta,\Omega,z)$, where
$\theta=K_\perp/k_0$ is the propagation angle with respect to the $z$ axis.

An accepted approach for understanding the structure of the CE relies upon the evaluation of the
MI gain profile of the PM modes of the NSE
\cite{mckinstrie:1992}. In Kerr self-focusing media with normal dispersion, the perturbations to a PM
mode that grow at maximum rate are those whose spatiotemporal frequencies $(K_\perp,\Omega)$ are related by
\begin{equation}\label{DISP1}
K_\perp(\Omega) =\sqrt{k_0k_0^{\prime\prime}\Omega^2 + 2k_0\tilde\beta}, \quad \mbox{($\tilde\beta >0$)},
\end{equation}
i.e., lie on a hyperbola on the $K_\perp$--$\Omega$ plane featuring an {\em angle gap} ($K_\perp$ gap)
\cite{mckinstrie:1992}. In this respect, Luther {\em et al.} proposed an intuitive picture that assumes the
largest MI gain to occur at angles and frequencies fulfilling the linear phase-matching condition of the
four-wave mixing (FWM) process supported by the Kerr response \cite{luther2:1994}. Under this
hypothesis, the asymptotic linear approximation $K_\perp(\Omega)\simeq \sqrt{k_0k_0^{\prime\prime}}|\Omega|$
was re-obtained, and the observed discrepancies were attributed to the
{\em nonlinear phase shift} produced by the PM mode {\em on the weak perturbation}.

\begin{figure}
\includegraphics[angle=0,width=7.5cm]{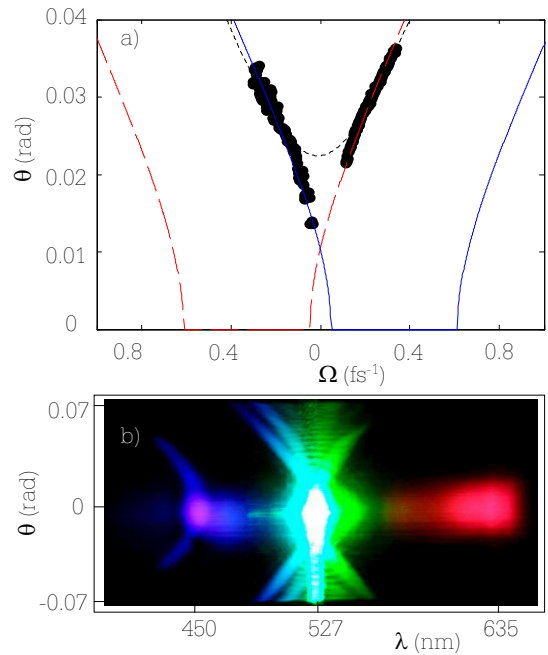}
\caption{\label{fig:fig2} (Color online) a) Closed circles: $\theta$--$\Omega$ distribution of peal fluence in Fig.
\ref{fig:fig1}; dotted line: best fit of the entire set of experimental data with Eq. (\ref{DISP1}); red-dashed
line: best fit of the $\Omega<0$ data with with Eq. (\ref{DISP}); blue-full line: best fit of the $\Omega>0$
data with with Eq. (\ref{DISP}) b) Experimentally measured $\theta$--$\lambda$ spectrum for
$E_{\rm in}=3$ $\mu$J, showing the Stokes and anti-Stokes sidebands.} %which according to the fit in (a)
%should occur at $635$ and $450$ nm, respectively.
\end{figure}

Preceding experimental observations of CE in borosilicate glass \cite{alfano:1970} and ethylene glycol
\cite{alfano:1993} were indeed interpreted to present a hyperbolic structure with an angle gap, which was
also attributed to the nonlinear phase shift. This trend is also visible in our measurements (Fig.
\ref{fig:fig1}) as an intersection of the two X arms at a non-zero angle.  If we attempt to fit
Eq. (\ref{DISP1}) (taking $\tilde\beta$ and $k_0^{\prime\prime}$ as free parameters) to the experimental
data [Fig.\ref{fig:fig2}(a), dotted line and closed circles, respectively] we obtain
$k^{\prime\prime}_0=0.053\pm 0.03$ fs$^{2}/\mu$m, which firmly supports the predictions
\cite{luther2:1994} regarding the asymptotic slope ($k^{\prime\prime}_0=0.055$ fs$^{2}/\mu$m at 527 nm
\cite{engen}); however, the angle-gap fitting strongly departs from experimental data in the non-asymptotic
region, which strongly calls for a fitting to a hyperbola with opposite curvature, that is, to a hyperbola
with a {\em frequency gap}. This observation advocates a novel interpretation of the origin of CE.

To understand our proposal, note first that the usual MI interpretation of the
seemingly angle-gap hyperbolic CE can be equivalently reread in terms of the
excitation of a weak, {\em linear X-wave mode} by a strong, nonlinear PM pump.
Indeed, Eq. (\ref{DISP1}) represents also the $K_\perp$--$\Omega$ spectrum of an
X-wave mode of the medium \cite{modos}, that is, a diffraction- and
dispersion-free pulsed Bessel beam in which cone-angle-induced dispersion
[$\theta(\Omega)\simeq K_\perp(\Omega)/k_0$] and material GVD balance mutually. The
spectrum in Eq. (\ref{DISP1}) belongs to a wave-mode of carrier frequency $\Omega_0$
and shortened axial wave vector $k_0-\tilde\beta$. In this reading, the excitation of
a X-wave mode with angle-gap ($\tilde\beta>0$) is a consequence of the nonlinear
phase matching between PM pump and X-wave for maximum efficiency of the interaction:
taking into account the nonlinear corrections to the wave vectors of both pump and
X-wave, phase matching imposes $k_0+\Delta k = k_0 -\tilde\beta + 2\Delta k$, where
$\Delta k= \omega_0n_2I/c$  is the {\em positive} nonlinear phase shift (in
self-focusing Kerr medium) for the PM pump of intensity $I$, and where $2\Delta k$ is
the corresponding nonlinear correction to the weak perturbation
\cite{alfano:1970,alfano:1993}, leading immediately to $\tilde\beta = \omega_0n_2I/c
>0$ (as predicted by the standard MI analysis \cite{mckinstrie:1992}).

We also note that a medium with GVD
[$k(\Omega)=k_0+k^{\prime}_0\Omega + k_0^{\prime\prime}\Omega^2/2$] supports more general X-wave modes with
strengthened and shortened wave vectors ($\tilde\beta$ negative and positive), as well as shifted carrier
frequency $\omega_0 + \tilde\Omega$, their most general $K_\perp$--$\Omega$ spectrum being expressed by
\cite{modos}
\begin{equation}\label{DISP}
K_\perp(\Omega) = \sqrt{k_0k_0^{\prime\prime}(\Omega-\tilde\Omega)^2 + 2k_0\tilde\beta}, \quad
\mbox{($\tilde\beta\gtreqqless 0$).}
\end{equation}
Their phase and group velocities are given by $v^{(p)}= (\omega_0+\tilde\Omega)/[k(\tilde\Omega)-\tilde\beta]$
and $v^{(g)}= 1/k'(\tilde\Omega)$, respectively, which can take both subluminal and superluminal values.
Equation (\ref{DISP}) describes a two-parameter family of hyperbolas, sharing the same asymptotic
slope $\sqrt{k_0k_0^{\prime\prime}}$, but admitting angle gaps ($\tilde\beta>0$) and frequency gaps
($\tilde\beta<0$), as well as positive and negative frequency shifts $\tilde\Omega$. X-wave spectra appear
then as a suitable tool for the description of the CE features.

The apparent frequency-gap hyperbolic form of the observed CE, and the actual absence of such a gap,
suggest that the CE cannot be described in terms of a single X-wave. This leads us to interpret the two
observed X tails as belonging to two different X waves both featuring frequency gaps, and proceed to fit Eq.
(\ref{DISP}) with two independent sets of parameters $\tilde\Omega$ and $\tilde\beta$ (and $k_0^{\prime\prime}
=0.056$ fs$^2/\mu$m) to the experimental data. This procedure also finds motivation in
the X-X structure of the numerically evaluated angular spectra of light filaments in water, which was
interpreted to be a consequence of pulse splitting on X-wave generation \cite{kolesikprl:2004}. Fittings
with $\tilde\Omega= +0.33$, $-0.33$ fs$^{-1}$ and $\tilde\beta= -2.2, -2.2\,$mm$^{-1}$ (dashed and continuous
lines, respectively) reproduce accurately the structure of the CE [Fig. \ref{fig:fig2}(a)]. The interpretation
in term of X-waves leads moreover to suspect the existence of two additional, not yet observed X arms in the
$K$--$\Omega$ spectrum at wavelengths 450 and 635 nm [Fig. \ref{fig:fig2}(a)].

To verify this prediction, we performed additional measurements with a lower resolution, 300 lines/mm
diffraction grating in order to cover a wider spectral region, as shown in Fig. \ref{fig:fig2}(b).
Alongside the central tails (around 527 nm), we clearly observe the generation of new frequencies in the
vicinity of 450 and and 635 nm, as expected. The reduced visibility of the red tail at 635 nm can be
attributed to the stronger linear absorption at this wavelength. In fact, we had to increase the input
pulse energy up to $E_{\rm in}= 3\,\mu$J in order to enhance the overall intensity of the new tails.

The question then arises of the mechanism responsible for the described spectral structure.
Here we propose that the double X-like CE is a result of a nonlinear FWM interaction ---the
lowest order process supported by the nonlinear Kerr response---, in which two signal and idler X-waves
experience a parametric amplification by a strong, {\em highly localized} (i.e., non PM) pump. If as
consistently with the self-focusing dynamics, we consider a strongly localized pump rather than a PM wave,
it is expected that nonlinear phase modulation does not influence significantly an extended object such an
X-wave mode, whose axial wave vector is solely determined by its cone angle, that is, by its weak,
longstanding tails far from the pump. In fact, in the X-wave the energy does not flow along the axial
direction, but along a conical surface, which prevents pump-induced cross-phase accumulation to take place.

Under this assumption, phase matching between a localized pump and a single X-wave at same frequency
$\omega_0$ would require $k_0+\Delta k=k_0-\tilde\beta$, a condition that contrarily to the case of a PM
pump, is satisfied by a frequency-gap X-wave mode ($\tilde\beta=-\Delta k<0$). More generally, if
$k=k_0 + \Delta k$ is the nonlinear pump wave number, and $k_{s,i}=
k(\tilde\Omega_{s,i})-\tilde\beta_{s,i}$ are two, signal and idler, X wave numbers, where
$k(\tilde\Omega_{s,i})=k_0+k'_0\Omega_{s,i}+ k_0^{\prime\prime}\Omega_{s,i}^2/2$, and  $\Omega_{s,i}$ are
their carrier frequency shifts, the conditions of energy and momentum conservation, $\tilde\Omega_s
=-\tilde\Omega_i$ and $2k=k_s+k_i$, for maximum efficiency of a FWM process involving two (degenerate)
highly localized pumps and the signal and idler X waves, leads to the relation $\tilde\beta_s +
\tilde\beta_i - k_0^{\prime\prime}\Omega_{s,i}^2= -2\Delta k$. Among all possible couples of X-wave modes
satisfying this relation, those whose spectra cross the pump, located around $\Omega=0, K_\perp=0$, are
the most energetically favored, since these X-waves will not need to grow from noise, but from the more
effective pump self-phase modulation. This condition was also found in \cite{contipre:2004} to result in the
largest gain in the case of a single X wave excited by a travelling pump. The X-wave--pump crossing
condition leads, from Eq. (\ref{DISP}), to $\tilde\beta_{s,i}= -k_0^{\prime\prime}\tilde\Omega_{s,i}^2/2$,
and then, on account that $\Delta k=\omega_0n_2I/c$, to
\begin{equation}\label{betaI}
\tilde\beta_{s,i}= - \omega_0 n_2 I/2c\,.
\end{equation}
Accordingly, the two X-waves will present symmetrically shifted carrier frequencies $\tilde\Omega_{i,s}=
\pm \sqrt{\omega_0n_2 I/ck_0^{\prime\prime}}$, and frequency-gaps of width $2|\tilde\Omega_{i,s}|$, from
$\Omega=0$ towards the Stokes and anti-Stokes bands, as observed in the experiment.

\begin{figure}[t]
\includegraphics[angle=-90,width=7.5cm]{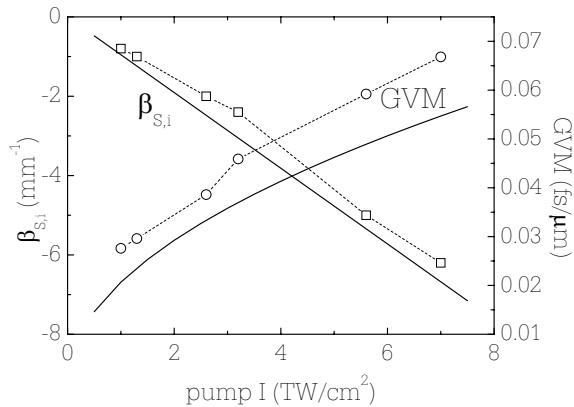}
\caption{\label{fig:fig3} Left: Values of $\tilde\beta_{s,i}$ obtained from X-wave fitting to the
numerical $K_\perp$--$\Omega$ spectra for increasing pump intensities $I$ (open squares), and
prediction of Eq. (\ref{betaI}) (solid curve). Right: GVM between the splitting pulses obtained from direct
space-time numerical simulation (open circles), and between the X-waves as given by
$2\sqrt{k_0^{\prime\prime}\omega_0 n_2 I/c}$ (solid curve).}
\end{figure}

Eq. (\ref{betaI}) predicts also a precise dependence of the whole CE structure on the
pump intensity $I$, whose validity we have tested numerically. We solved the NSE
(\ref{eq:NLSE}) with $k_0^{\prime\prime}= 0.056$ fs$^2/\mu$m, $n=1.33$ and $n_2=
1.6\times 10^{-16}$ cm$^2/$W, for an input 200 fs-long, 100 $\mu$m-wide (FWHM)
Gaussian wave packet. Since CE is seen to develop explosively at the collapse,
we identified the pump intensity $I$ with
the absolute peak intensity reached during propagation. In order to attain different
values of this intensity without changing the rest of parameters, we added to the
second member of the NSE (\ref{eq:NLSE}) the nonlinear dissipation term
$-\beta^{(K)}|A|^K A/2$, with $K=3$ (three-photon absorption at 527 nm in water), and
with $\beta^{(K)}$ ranging from $1.2\times 10^{-23}$ to $8\times 10^{-25}$
cm$^3$/W$^2$. The $K_\perp-\Omega$ spectra obtained numerically are symmetric with
respect to $\Omega=0$ due to the approximations involved in the NSE, and present a
nearly invariant X-X structure beyond the collapse point (in spite of the quickly
decreasing peak intensity $I$), with one tail of each X wave nearly passing through
$\Omega=0$, as observed experimentally. The spectrum at $3$ cm from the input plane
was then easily fitted with two, signal and idler, material X-waves modes crossing
$\Omega=0$ [i.e., with $k_0^{\prime\prime}=0.056$ fs$^2/\mu$m, and with
$\tilde\Omega_{s,i}=\pm (-2\tilde\beta_{s,i}/k_0^{\prime\prime})^{1/2}$],
$\tilde\beta_{s,i}$ being then the only one free parameter. Figure \ref{fig:fig3}
shows the values of $\tilde\beta_{s,i}$, obtained from the best fits to the spectra,
versus pump intensity $I$ (open squares). The excellent agreement with the predicted
$\tilde\beta_{s,i}$--$I$ dependence in Eq. (\ref{betaI}) (solid line) strongly
supports the FWM analysis.

Extending further our interpretation, we conjecture that the well-known phenomenon of
pulse temporal splitting in filamentation with normal GVD, usually described from a purely temporal
perspective, is a consequence of higher-order matching among the interacting waves ---in our case, of
{\em group matching}. Indeed,
the zero-order, phase-matching condition (\ref{betaI}) entails that the two X-waves must travel at
different group velocities $v^{(g)}_{s,i} = 1/k'(\Omega_{s,i})= 1/[k'_0 + k_0^{\prime\prime}\Omega_{s,i}]$,
and therefore split apart in time one from another with a group velocity mismatch (GVM)
$1/v^{(g)}_s - 1/v^{(g)}_{i}= 2 k_0^{\prime\prime}|\tilde\Omega_{s,i}| = 2\sqrt{k_0^{\prime\prime}
\omega_0n_2I/c}$ proportional to the square root of the pump intensity. Group matching among the
interacting waves is then better attained if the two pump waves, breaking their initial degeneracy, split
also to co-propagate with the X-waves. In this view, pulse splitting is not a mere
collapse-arresting mechanism \cite{luther2:1994}, which further determines, taken it for granted, the
generation of two X-waves \cite{kolesikprl:2004}; instead, pulse splitting emerges here as a particular
feature of the phase and group matched wave configuration most favored by the FWM nonlinear interaction
inherent to the Kerr NSE dynamics. In other words, X-wave instability or parametric amplification, and
splitting instability are not independent phenomena, but two aspects of the same process. To sustain this
hypothesis, we have evaluated the GVM of the split pumps in the same numerical simulations as in the
preceding paragraph. To do so, we plotted the temporal profiles at different propagation distances
beyond collapse, and obtained the GVM as the slope of a linear fitting to the
delay between the two temporal peaks versus propagation distance. Figure \ref{fig:fig3} shows that the
pump GVM (open circles) does not depart more than 10 percent from the predicted X-wave GVM
(solid curve), and that both follow a similar dependence with the square root of the intensity $I$
at collapse.

In summary, our experiments demonstrate that CE emission in the angular spectrum of filaments is not
interpretable in the frame of MI of PM modes of the NSE. Strong localization of the self-focusing pulse
substantially modifies the MI pattern, which finds accurate description in terms of linear X-wave modes of
the medium, and simple explanation as a result of a dominant, phase-matched FWM mixing process supported by
the NSE dynamics between two highly localized, strong pump waves and two amplifying weak X-waves. Pulse
temporal splitting emerges in this model as the necessary temporal dynamics for preserving group matching
among the interacting waves.

%\bibliography{main_bibliography}

\end{document}